# Optimization of Energy Efficient Transmission in Underwater Sensor Networks

Hongkun Yang     Bin Liu

**Abstract**

Underwater communication is a challenging topic due to its singular channel characteristics. Most protocols used in terrestrial wireless communication can not be directly applied in the underwater world. In this paper, we focus on the issue of energy efficient transmission in underwater sensor networks (UWSNs) and analyze this problem in a rigorous and theoretical way. We formalize an optimization problem which aims to minimize energy consumption and simultaneously accounts for other performance metrics such as the data reliability and the communication delay. With the help of Karush-Kuhn-Tucker conditions (KKT conditions), we derive a simple and explicit, but nevertheless accurate, approximate solution under reasonable assumptions. This approximate solution provides theoretical guidelines for designing durable and reliable UWSNs. Our result also shows that reliability and communication delay are crucial factors to the energy consumption for transmission.

**Energy efficient transmission, underwater sensor networks, KKT conditions.**

## 1 Introduction

Understanding the key mechanisms of the oceans is crucial to the knowledge of our Earth's climate and atmosphere. Over the past few years, there has been a relentless effort to investigate this underwater world which remains little explored. This infatuation is highly motivated by various commercial, scientific, and military applications. The underwater sensor network (UWSN), a promising technique for the human being to explore the oceans, has a myriad of applications such as mineral exploitation, environmental monitoring, disaster prevention, military surveillance and coastline protection.

Communication in underwater is a very challenging topic which is featured by large propagation delay, limited available bandwidth, high error probability and severe energy limitation. These features make the design of transmission mechanisms rather awkward. Among these features, energy efficiency is critical for a UWSN to maximize its utility.

UWSNs are typically characterized by severely energy-limited nodes. For one thing, the nodes have small batteries and therefore cannot afford much energy to complete tasks. For another, their power supplies cannot typically be replenished; that is, once exhausted, the node is discarded. Thus, efficient use of energy is of great importance to long-sustained and well-operated UWSNs. In this paper, we focus on the issue of energy efficient transmission.

The issue of energy efficiency is widely considered in the literature. Various data transmission mechanisms [1–6] are proposed to achieve the objective of energy efficiency. These efforts are of great value to reduce energy consumption in UWSNs. However these mechanisms are designed more or less based on heuristics, and a theoretical analysis of energy consumption in UWSNs is little addressed.

In this paper, we analyze energy efficient transmission in a rigorous and theoretical way. We formalize an optimization problem which aims to minimize energy consumption but also comprehensively considers other performance metrics such as data reliability and communication delay. Using this problem, we study the effect of four key parameters (node distance, communication frequency, packet length and SNR) on the performance of UWSNs. Utilizing a two-step approach, we simplify this problem, and derive an explicit approximate solution to the simplified problem according to Karush-Kuhn-Tucker conditions (KKT conditions) [7]. We show that the approximate solution is accurate with regard to the simplified problem, and it also achieves near optimum of the original optimization problem. This approximate solution provides theoretical guidelines for designing durable and reliable UWSNs. Our result also shows that reliability and communication delay are crucial factors to the energy consumption for transmission.

The rest of the paper is organized as follows. Section 2 introduces the system model of our paper. Section 3 formalizes the optimization problem and reduces it to a simplified version. Section 4 solves the simplified problem with the help of KKT conditions. Numerical result is presented in Section 5. Finally conclusions are drawn in Section 6.

## 2 Preliminaries

### 2.1 Scenario

In this paper, we consider the scenario as is depicted in Fig. 1. There is a source node and a destination node. And the distance between the source and the destination is $d$. The source node keeps sending a package repeatedly until the destination node receives it correctly. This scenario, simple though it is, can be regarded as a basic component of multi-hop scenarios. Moreover, by studying it thoroughly, we can get valuable insights into more sophisticated schemes such as Forward-Error-Correcting based data transport protocols.

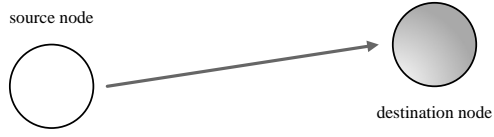

Figure 1: The network scheme.

## 2.2 Energy Model

The passive sonar equation [8] indicates that the signal to noise ratio (SNR) per bit $\gamma_b$ of an emitted underwater signal at the receiver

$$\gamma_b = SL - TL - NL + DI \qquad (1)$$

where $SL$ is the source level, $TL$ the transmission loss, $NL$ the noise level and $DI$ the directivity index. The unit of all quantities is dB. Since omnidirectional hydrophones are used, $DI$ is 0.

The transmission loss $TL$ over a distance $d$ (in m) for a signal of frequency $f$ (in kHz) is

$$TL = 10 \lg d + \alpha(f) \times d \times 10^{-3} \qquad (2)$$

where $\alpha(f)$ is the absorption coefficient in dB/km. It follows from the Thorp's formula [9] that

$$\alpha(f) = \frac{0.11 f^2}{1 + f^2} + \frac{44 f^2}{4100 + f^2} + 2.75 \times 10^{-4} f^2 + 0.003 \qquad (3)$$

The noise level $NL$ is affected by four sources: turbulence, shipping, waves, and thermal noise. For the sake of convenience, we simplify $NL$ with a practical approximation [10]

$$NL = 50 - 18 \lg f \qquad (4)$$

Combining (1), (2), (3) and (4), we obtain

$$SL = 50 + \gamma_b + 10 \lg d + \alpha(f) \times d \times 10^{-3} - 18 \lg f \qquad (5)$$

On the other hand, $I_t$ (W/m$^2$), the transmitted signal intensity at 1m from the source, is expressed as follows

$$I_t = 10^{SL/10} \times 0.67 \times 10^{-18} \qquad (6)$$

Thus, the transmitter power $P_t$, which achieves intensity $I_t$ at a distance of 1m from the source in the direction of the receiver, is

$$P_t = 2\pi \times H \times I_t \qquad (7)$$

where $P_t$ is in W, and $H$ is the water depth in m.

Utilizing (1)-(7), we can express $\gamma_b$ with $P_t$, $d$, and $f$

$$\begin{aligned}\gamma_b =& 10 \times (\lg P_t - \lg(2\pi H \times 0.67 \times 10^{-18}) - \lg d) \\ & - \alpha(f)\, d \times 10^{-3} + 18 \lg f \end{aligned} \qquad (8)$$

## 2.3 Channel Model

In our model, we employ binary phase shift keying (BPSK), one of the most commonly used modulation schemes. In BPSK, the SNR per symbol, $\gamma_s$, equals the SNR per bit, $\gamma_b$, that is, $\gamma_s = \gamma_b$.

The multipath effect is distinct in the underwater channel, so a Rayleigh fading channel is an appropriate model for the underwater channel (both in shallow and deep water). The average BER of BPSK in an underwater Rayleigh fading channel is [11]

$$P_b(\gamma_b) = \frac{1}{2}(1 - \sqrt{\frac{\overline{\gamma}_s}{1+\overline{\gamma}_s}}) \tag{9}$$

where $\overline{\gamma}_s = 10^{\gamma_s/10} = 10^{\gamma_b/10}$.

# 3 Problem Formulation

We concentrate on the energy consumption and its relations to the four key parameters (node distance, communication frequency, packet length, and transmitter power) in UWSNs. We also take into account other performance metrics such as reliability and communication delay.

We optimize the energy consumption through the metric $E_b$, average energy consumption for successfully transmitting one bit payload. $E_b$ can be written as follows

$$E_b = \frac{E_{per}}{P_{acc} \cdot L_p} \tag{10}$$

where $E_{per}$ is the energy consumption per transmission attempt, $P_{acc}$ is the packet acceptance ratio, and $L_p$ is the length of the payload for one packet. Meanwhile, some constraints should be satisfied. First, the packet acceptance ratio $P_{acc}$ should be no less than a threshold $P_{acc0}$ to maintain the reliability. In addition, a packet must contain payload, and the communication frequency $f$ and the transmitter power $P_t$ must be nonnegative.

Thus, we come to an optimization problem

$$\min_{P_t, L, f} E_b = \frac{E_{per}}{P_{acc} \cdot L_p}$$
$$\text{subject to } \begin{cases} P_t \geq 0 \\ L \geq \mu + \tau + 1 \\ P_{acc} \geq P_{acc0} \\ f \geq 0 \end{cases} \tag{11}$$

where $L$ is the length of a packet, $\mu$ the length of the header, and $\tau$ the length of the trailer of a packet. Thus $L_p = L - (\mu + \tau)$. The optimization problem (10) aims to minimize the energy consumption and maintains the reliability at the same time. Moreover, the constraint of $P_{acc}$ is also responsible for the communication delay. The link delay takes a dominating part in the total time cost for one transmission (we will discuss it in Section 5). For a given distance $d$, the link delay between the source node and the destination node of one transmission attempt is fixed. Thus the total time cost for successfully transmitting one packet, which is mainly composed of the link delay, largely depends on the packet acceptance ratio $P_{acc}$.

Now we develop an expression of $E_b$ using $P_t$, $L$ and $f$. We have known that

$$L_p = L - \mu - \tau \tag{12}$$

$P_{acc}$ can be expressed as

$$P_{acc} = (1 - P_b(\gamma_b))^L = \left(0.5 + 0.5\sqrt{\frac{10^{\gamma_b/10}}{1 + 10^{\gamma_b/10}}}\right)^L \tag{13}$$

where $\gamma_b$ can be calculated using (8) in terms of $P_t$, $d$ and $f$.

Furthermore, we also assume that if the bandwidth is $f$ kHz, the available bit rate is $f$ kb/s. In addition, for commercial hydrophones, the received energy of each packet is typically around one fifth of the transmitted energy $P_t$ [12]. Besides, a sensor node also consumes some extra power each time it transmits or receives a packet, such as energy used by MPU/DPS, A/D or D/A converter, filter and so on. This energy consumption should also be considered and we denote it by $P_c$. So the total energy consumption for one transmission attempt is

$$E_{per} = \frac{L}{1000f} \times (\frac{6}{5}P_t + P_c) \tag{14}$$

where $L$ is the packet length.

### 3.1 A Simplification of (11)

From (12)-(14), we find that in the optimization problem (11), the terms which involve $f$ are rather complicated. This incurs great difficulties in solving (11). To reduce complexity, we adopt a two-step approach to analyze (10). We first find optimal communication frequency $f_*(d)$ which minimizes the transmission power $P_t$ with respect to $d$, and reduce (10) to an optimization problem of two variables ($P_t$ and $L$). Second, we solve the reduced optimization problem with the help of KKT conditions.

From (6) and (7), we know that $SL$ directly determines the transmission power $P_t$ and that minimizing $P_t$ is essentially minimizing $SL$. So we differentiate (5) with respect to $f$ and obtain

$$\begin{aligned}\frac{\partial SL}{\partial f} =& \Big(\frac{2.2 \times 10^{-4}}{(1+f^2)^2} + \frac{360.8}{(4100+f^2)^2} \\ &+ 5.5 \times 10^{-7}\Big)fd - \frac{18}{f \ln 10}\end{aligned} \tag{15}$$

The node distance $d$ in UWSNs is generally within the range of 0.1km and 100km [13]. Moreover, it can be shown that for $0.1\text{km} \leq d \leq 100\text{km}$

$$\lim_{f \to 0^+} \frac{\partial SL}{\partial f} = -\infty, \text{ and } \lim_{f \to +\infty} \frac{\partial SL}{\partial f} = +\infty. \tag{16}$$

Thus it can be reasoned that $\forall d \in [0.1, 100]$, there exists an optimal frequency $f_*(d)$ which minimizes $SL$ and hence $P_t$, and that $\frac{\partial SL}{\partial f}\big|_{f=f_*(d)} = 0$. So we can find $f_*(d)$ by letting $\frac{\partial SL}{\partial f} = 0$ and numerically solving the equation. The relation between $f_*(d)$ and $d$ is illustrated in Fig. 2.

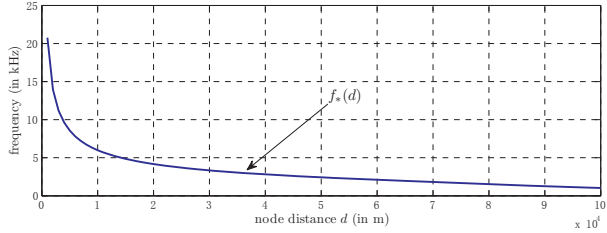

Figure 2: $f_*(d)$ vs. $d$.

Substituting $f_*(d)$ for $d$, we can reduce (11) to a problem with two variables which has a relatively plain form

$$\min_{P_t, L} E'_b = \frac{C_2 L(5P_c + 6P_t)}{(L - \mu - \tau)\left(0.5 + 0.5\sqrt{\frac{C_1 P_t}{C_0 + C_1 P_t}}\right)^L}$$

$$\text{subject to } \begin{cases} P_t \geq 0 \\ L \geq \mu + \tau + 1 \\ P_{acc} \geq P_{acc0} \end{cases} \tag{17}$$

where $C_0 = 4.10 \times 10^{-13} \times 10^{10^{-4} d \times \alpha(f_*(d))} dH$, $C_1 = f_*(d)^{1.8}$ and $C_2 = \frac{1}{5000 f_*(d)}$. Note that $C_0$, $C_1$ and $C_2$ vary with $d$, and they can be regarded as constants for a fixed $d$.

For the convenience of further mathematical manipulation, we use the natural logarithm of $E'_b$ as the objective function instead of $E'_b$, which does not change the optimal solution. Thus, we obtain

$$\min_{P_t, L} \ln E'_b = \ln C_2 + \ln L + \ln(5P_c + 6P_t) - \ln(L - \mu - \tau)$$

$$- L \ln(0.5 + 0.5\sqrt{\frac{C_1 P_t}{C_0 + C_1 P_t}}) \text{ subject to } \begin{cases} L \geq \mu + \tau + 1 \\ P_{acc} \geq P_{acc0} \\ P_t \geq 0 \end{cases} \tag{18}$$

After exploring qualitative properties of (18), we will solve this problem using KKT conditions. In our model, we set $H = 10\text{m}$, $P_c = 1\mu\text{W}$, and $\mu = \tau = 16\text{bits}$.

# 4 Finding Minimal Energy Consumption

## 4.1 An Analysis of the Constraints

The feasible region of (18) is determined by the three inequalities in (18). However, we will show that one of them is redundant, so they can be further simplified.

Note that
$$P_{acc} = \left(0.5 + 0.5\sqrt{\frac{C_1 P_t}{C_0 + C_1 P_t}}\right)^L. \tag{19}$$

We let $0.5 < P_{acc0} < 1$, which is a reasonable assumption, so the inequality $P_{acc} \geq P_{acc0}$ can be rewritten as

$$P_t \geq \frac{C_0(-1 + 2P_{acc0}^{\frac{1}{L}})^2}{4C_1 P_{acc0}^{\frac{1}{L}}(1 - P_{acc0}^{\frac{1}{L}})} \geq 0 \tag{20}$$

Thus, we can remove $P_t \geq 0$ from the set of constraints without changing the feasible region of the original optimization problem.

According to (20), we obtain an equivalent set of constraints

$$\begin{cases} P_t - \dfrac{C_0(-1 + 2P_{acc0}^{\frac{1}{L}})^2}{4C_1 P_{acc0}^{\frac{1}{L}}(1 - P_{acc0}^{\frac{1}{L}})} \geq 0 \\ L - (\mu + \tau + 1) \geq 0 \end{cases} \tag{21}$$

We denote the first inequality by $h_1(P_t, L)$ and the second inequality by $h_2(P_t, L)$. Then the optimization problem (18) is transformed to an equivalent one which has two inequality constraints

$$\min_{P_t, L \in R} \ln E'_b \quad \text{subject to} \quad \begin{cases} h_1(P_t, L) \geq 0 \\ h_2(P_t, L) \geq 0 \end{cases} \tag{22}$$

An illustration of the feasible region of (22) can be found in Fig. 3 (the hatched area).

## 4.2 Qualitative Properties of (22)

Before solving this problem, we first explore some of its qualitative properties.

First of all, we explain some terminologies with respect to (22). The *feasible region* is the set of points which satisfy all the constraints. We say that an inequality constraint $h_i(x) \geq 0$ is *effective* at a point $x = (P_t, L)$ if the constraint holds with equality at $x$, that is, we have $h_i(x) = 0$. $E$ denotes the set of effective constraints at $x$. Obviously, $E \subseteq \{h_1, h_2\}$. If $D(E) = \{\nabla h_i(x), h_i \in E\}$ is linearly independent, we say that the *constraint qualification* is met at $x$. Note that $\nabla$ is the gradient operator. And if $E = \emptyset$, we say that the constraint qualification also holds.

The following two lemmas are of great value for finding the optimal solution to (22).

**Lemma 1.** *The optimization problem (22) satisfies the constraint qualification at any $x = (P_t, L)$ within the feasible region.*

*Proof.* Note that

$$\begin{cases} \nabla h_1(x) = (1, \dfrac{C_0 P_{acc0}^{-1/L} \ln P_{acc0} \left(-1 + 2P_{acc0}^{\frac{1}{L}}\right)}{4L^2 C_1 \left(-1 + P_{acc0}^{\frac{1}{L}}\right)^2})^T \\ \nabla h_2(x) = (0, 1)^T \end{cases}$$

and that $D(E) \subseteq \{\nabla h_1(x), \nabla h_2(x)\}$. If $D(E) \neq \emptyset$, it is obvious that $D(E)$ is linearly independent, so the constraint qualification holds. If $D(E) = \emptyset$, since the constraint qualification pertains only to effective constraints, it holds vacuously in this case also. □

**Lemma 2.** *A global minimum exists to the optimization problem (22).*

*Proof.* In the expression of $\ln E'_b$, we have $\ln L > \ln(L - \mu - \tau), \forall L \geq \mu + \tau + 1$, and $\ln(0.5 + 0.5\sqrt{\frac{C_1 P_t}{C_0 + C_1 P_t}}) < 0, \forall P_t \geq 0$, so we have

$$\ln E'_b \geq \ln(5P_c + 6P_t) + \ln C_2 \qquad (23)$$

For a given $(P_{t0}, L_0)$ in the feasible region, there exists an $M_{P_t}$ such that $M_{P_t} > P_{t0}$ and that $\forall P_t > M_{P_t}, \ln(5P_c + 6P_t) + \ln C_2 > \ln E'_b(P_{t0}, L_0)$. Using (23), we know that $\ln E'_b(P_t, L) \geq \ln(5P_c + 6P_t) + \ln C_2 > \ln E'_b(P_{t0}, L_0), \forall P_t > M_{P_t}$.

Thus, the feasible region of (22) can be divided into two areas by $M_{P_t}$, as is shown in Fig. 3. Note that $(P_{t0}, L_0)$ lies within Area I. Moreover, according to the previous analysis, we know that $\forall (P_t, L)$ which lies within Area II, $\ln E'_b(P_t, L) > \ln E'_b(P_{t0}, L_0)$.

On the other hand, Area I with its border is bounded and closed, so it is compact and the Weierstrass Theorem guarantees that $\ln E'_b$ has a minimum $(P_t^*, L^*)$ in Area I with its border. Given the fact $(P_{t0}, L_0)$ lies within Area I, we can see that $\ln E'_b(P_{t0}, L_0) \geq \ln E'_b(P_t^*, L^*)$. Since $\ln E'_b(P_t, L) > \ln E'_b(P_{t0}, L_0), \forall (P_t, L)$ which lies within Area II, we conclude convincingly that $(P_t^*, L^*)$ is a global minimum to the optimization problem (22). □

After proving the above two lemmas, we introduce the following theorem, which claims the validity of our method.

**Theorem 1.** *Suppose the following conditions hold:*

1. *A global minimum $x^* = (P_t^*, L^*)$ exists to the optimization problem (22).*

2. *The constraint qualification is met at $x^*$.*

*Then, the global minimum $x^*$ is a solution to KKT conditions of (22).*

*Proof.* Due to a lack of space, please refer to [7]. □

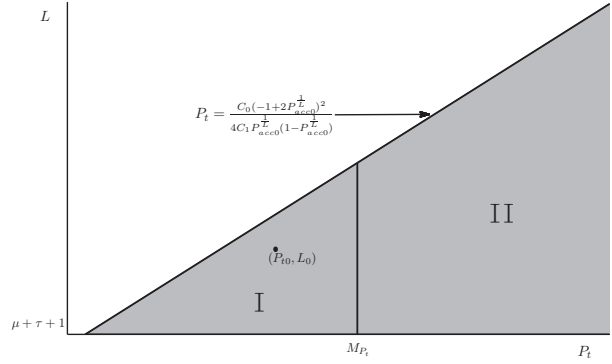

Figure 3: The feasible region (the hatched area) of (22).

## 4.3 Finding the optimal solution using KKT conditions

We have shown that the optimization problem (22) has a global minimum and that the constraint qualification is met at the minimum. According to Theorem 1, we can find the optimal solution by solving KKT conditions. The KKT conditions [7] of (22) are listed as follows.

$$\frac{\partial \ln E'_b}{\partial P_t} - \lambda_1 \frac{\partial h_1}{\partial P_t} - \lambda_2 \frac{\partial h_2}{\partial P_t} = 0 \tag{24a}$$

$$\frac{\partial \ln E'_b}{\partial L} - \lambda_1 \frac{\partial h_1}{\partial L} - \lambda_2 \frac{\partial h_2}{\partial L} = 0 \tag{24b}$$

$$h_1(P_t, L) \geq 0 \tag{24c}$$

$$h_2(P_t, L) \geq 0 \tag{24d}$$

$$\lambda_1 h_1(P_t, L) = 0 \tag{24e}$$

$$\lambda_2 h_2(P_t, L) = 0 \tag{24f}$$

$$\lambda_1, \lambda_2 \geq 0 \tag{24g}$$

Equations (24a)-(24g) are solved through a case by case analysis.

### 4.3.1 Case 1, $\lambda_1 \neq 0, \lambda_2 \neq 0$

In this case, (24e) and (24f) indicate that $h_1(P_t, L) = 0, h_2(P_t, L) = 0$. Thus,

$$\begin{cases} P_t = \dfrac{C_0 P_{acc0}^{-\frac{1}{1+\mu+\tau}} \left(1 - 2P_{acc0}^{\frac{1}{1+\mu+\tau}}\right)^2}{4C_1 \left(1 - P_{acc0}^{\frac{1}{1+\mu+\tau}}\right)} \\ L = \mu + \tau + 1 \end{cases} \tag{25}$$

The objective function

$$\ln E'_b = \ln C_2 + \ln\left(5P_c - \frac{3C_0(1-2Y)^2}{2C_1(-1+Y)Y}\right) \quad (26)$$
$$+ \ln(Y^{-1-\mu-\tau}) + \ln(1+\mu+\tau)$$

where $Y = P_{acc0}^{\frac{1}{1+\mu+\tau}}$.

In this case, $(P_t, L)$ obtained from (25) is already a feasible solution to (22).

**4.3.2 Case 2, $\lambda_1 \neq 0, \lambda_2 = 0$**

In this case, (24e) requires that

$$P_t = \frac{C_0(-1 + 2P_{acc0}^{\frac{1}{L}})^2}{4C_1 P_{acc0}^{\frac{1}{L}}(1 - P_{acc0}^{\frac{1}{L}})}, \quad (27)$$

that is

$$P_{acc0} = P_{acc} = \left(0.5 + 0.5\sqrt{\frac{C_1 P_t}{C_0 + C_1 P_t}}\right)^L \quad (28)$$

Combining (27), (24a) and (24b), we can obtain an equation of $X$ ($X = \sqrt{\frac{C_1 P_t}{C_1 P_t + C_0}}$) whose intricate appearance makes its solution a formidable task. So we follow an intuitive and pellucid way to tackle this problem. Observing that $Pacc0 \approx 1$ and $\frac{1}{L} \approx 0$, we approximate $Pacc0^{\frac{1}{L}}$ to

$$Pacc0^{\frac{1}{L}} \approx 1 + \frac{\ln(Pacc0)}{L} \quad (29)$$

Substitute (29) into (27), we get

$$P_t = -\frac{C_0}{4C_1}\left(\frac{\ln(P_{acc0})}{L} + \frac{L}{\ln(P_{acc0})} + 3\right) \quad (30)$$

Here it is more convenient to take $E'_b$ as the objective function. So we apply (29) and (28) to the expression of $E'_b$ in (17), and get

$$E'_b = \frac{C_2}{P_{acc0}} \frac{L(5P_c + 6P_t)}{L - \mu - \tau}$$
$$= \frac{C_2}{P_{acc0}} \left(-\frac{1}{2C_1 \ln P_{acc0}}(\frac{A}{L_p} + 3C_0 L_p) + B\right) \quad (31)$$

where $L_p = L - \mu - \tau$. In (31),

$$A = \ln P_{acc0}(\mu + \tau)(9C_0 - 10C_1 P_c) + 3C_0(\ln P_{acc0})^2$$
$$+ 3C_0(\mu + \tau)^2$$
$$B = \frac{\ln P_{acc0}(10C_1 P_c - 9C_0) - 6C_0(\mu + \tau)}{2C_1 \ln P_{acc0}}$$

It can be showed that $\ln P_{acc0} < 0$ and $A > 0$, so when

$$L = \mu + \tau + \sqrt{\frac{A}{3C_0}} \tag{32}$$

$E'_b$, or $\ln E'_b$, achieves its minimum. The corresponding optimal $P_t$ can be computed using (30).

In this case, we attain an explicit approximation instead of an exact answer. In Section 5, we will show that this approximation receives satisfactory results. More specifically, given a wide range of $d$ and a practical value of $P_{acc0}$, the $(P_t, L)$ obtained from (30) and (32) can meet the threshold packet acceptance ratio $P_{acc0}$ and save considerable energy.

### 4.3.3  Case 3, $\lambda_1 = 0, \lambda_2 \neq 0$

In this case, $L = \mu + \tau + 1$ and $\frac{\partial \ln E'_b}{\partial P_t} = 0$. Applying $L = \mu + \tau + 1$ to (24a) and simplifying it, we obtain

$$a_0 + a_1 X + a_2 X^2 + a_3 X^3 = 0 \tag{33}$$

where $X = \sqrt{\frac{C_1 P_t}{C_0 + C_1 P_t}}$ and $X \in (0, 1)$. The coefficients of (33) are

$$a_0 = 5C_1 P_c L \qquad\qquad a_1 = -12C_0 - 5C_1 P_c L$$
$$a_2 = (6C_0 - 5C_1 P_c)L \qquad\qquad a_3 = -(6C_0 - 5C_1 P_c)L$$

We denote the left part of (33) by $T_3(X)$. Note that $T_3(0^+) = 5C_1 P_c(1+\mu+\tau) > 0$ and that $T_3(1^-) = -12C_0 < 0$. Thus (33) has a root within $(0,1)$ which can be found by numerical solvers such as Bisection method.

### 4.3.4  Case 4, $\lambda_1 = 0, \lambda_2 = 0$

Combining (24a) and (24b), we obtain

$$f_1(X) + f_2(X) - \frac{\mu + \tau}{2} = 0 \tag{34}$$

where

$$f_1(X) = \frac{1}{2} \frac{24 C_0 X}{(-1+X)(-6 C_0 X^2 + 5 C_1 P_c(-1+X^2))}$$

$$f_2(X) = \frac{\sqrt{(\mu+\tau)(\ln 16 + (\mu+\tau)(\ln(\frac{1+X}{2}))^2 - 4\ln(1+X))}}{2 \ln\left(\frac{1+X}{2}\right)}$$

$X = \sqrt{\frac{C_1 P_t}{C_0 + C_1 P_t}}$ and $X \in (0, 1)$. We denote the left part of (34) by $T_4(X)$. Note that $T_4(0^+) < 0$ and that $T_4(1^-) = +\infty$. Thus, (34) has a root within $(0,1)$.

## 5 Numerical Result

In this section, we assess the solutions to the four cases in Section 4, and obtain the conclusion that (30) and (32) constitute an accurate approximate solution to (22) given a considerably wide range of $d$ and a practical value of $P_{acc0}$. We also show that our two-step approach to analyze (11) can effectively find near optimum of (11).

### 5.1 Feasibility of Solutions to (22)

In Section 4, we obtain four solutions (four cases) to (22). Here, we check whether these solutions satisfy two constraints ($h_1$ and $h_2$) of (22). It can be shown that for each case, the constraint of $L$ ($h_2$) is met. Thus we investigate the packet acceptance ratio $P_{acc}$ of each solution.

For Case 3 and Case 4, $\lambda_1 = 0$, so the solutions are independent of $P_{acc0}$ and $h_1$ might not be effective in Case 3 and Case 4. Fig. 4 depicts the packet acceptance ratio of solutions to Case 3 and Case 4. We can see that in both cases, $P_{acc}$ decays fast as the node distance $d$ grows. On the other hand, a typical threshold $P_{acc0}$ is usually no less than 95%. So the solutions to these two cases are not feasible except for very small node distance. This indicates that reliability and communication delay are crucial factors to the energy efficient transmission in UWSNs. In other words, considering energy consumption alone incurs high packet loss and large communication delay.

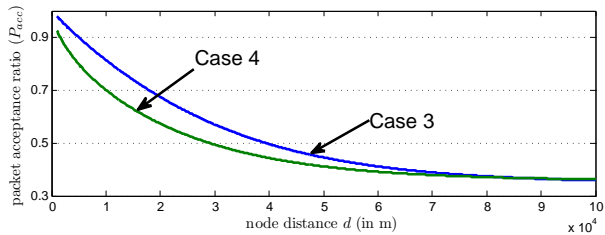

Figure 4: Packet acceptance ratio of solutions to Case 3 and Case 4.

According to (25), the solution to Case 1 is a feasible solution to (22) (Fig. 5). Although in Case 2, we derive an approximate solution instead of an exact solution, we find in Fig. 5 that $P_{acc}$ of the approximate solution converges to $P_{acc0}$ fast as $d$ grows. In addition, for a higher $P_{acc0}$, the converge becomes faster. In other words, the approximate solution to Case 2 satisfies $h_1$ for a sufficiently large $d$ and a sufficiently high $P_{acc0}$.

### 5.2 Optimality of the Approximate Solution

According to the above analysis, we find that the solution to Case 1, and the approximate solution to Case 2 are feasible with respect to $h_1$ and $h_2$. In Fig.

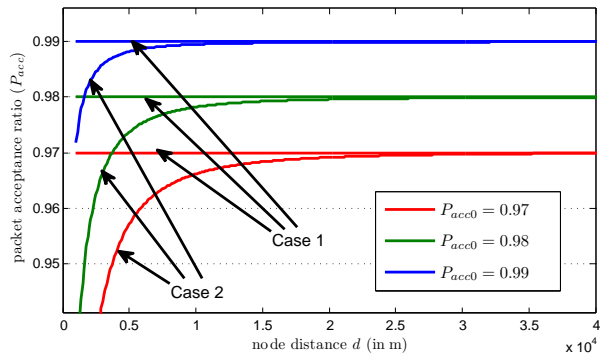

Figure 5: Packet acceptance ratio of solutions to Case 1 and Case 2 with respect to different $P_{acc0}$.

6, we see that the value of objective function corresponding to the solution to Case 1 lies in the upper part of the graph. Thus the approximate solution attains better performance than the solution to Case 1. We also numerically solve (22), and compare the numerical solution with the approximate solution (Table 1). We convincingly conclude that the approximate solution to Case 2 is an accurate approximation of the solution to (22).

| $P_{acc0}$ \ $d$ | 10000 | 20000 | 30000 | 40000 | 50000 |
|---|---|---|---|---|---|
| 0.980 | 0.012 | 0.015 | 0.015 | 0.015 | 0.015 |
| 0.985 | 0.009 | 0.011 | 0.011 | 0.012 | 0.012 |
| 0.990 | 0.007 | 0.000 | 0.000 | 0.000 | 0.000 |
| $P_{acc0}$ \ $d$ | 60000 | 70000 | 80000 | 90000 | 100000 |
| 0.980 | 0.015 | 0.015 | 0.015 | 0.015 | 0.015 |
| 0.985 | 0.012 | 0.012 | 0.012 | 0.012 | 0.012 |
| 0.990 | 0.000 | 0.000 | 0.000 | 0.000 | 0.000 |

Table 1: Relative error (%) of the approximate solution compared to the numerical solution to (22)

In Section 4, we adopt a two-step approach to analyze the original optimization problem (11). Now we numerically solve (11), and compare the numerical solution with the approximate solution in Fig. 6. We see that the approximate solution we derive can find near optimum of the original problem (11).

### 5.3 Verification of the Assumption in Section 3

We verify the assumption made in Section 3 which states that the link delay takes a dominating part in the total time cost for one transmission. We denote

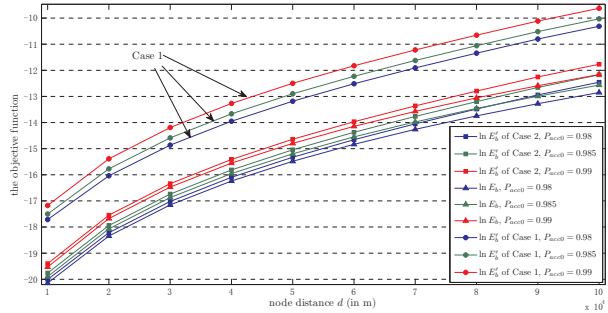

Figure 6: Comparison of the objective functions of different solutions. The circle marked data represents the solution to Case 1, the square marked data represents the approximate solution to Case 2, and the triangle marked data represents the numerical optimal solution to (11). Different colors correspond to different $P_{acc0}$.

the link delay by $T_1$ and

$$T_1 = \frac{d}{v}$$

where $d$ is the node distance and $v$ is the propagation speed of acoustic signals in water (here we let $v = 1500$m/s). On the other hand, the sending or receiving time for a packet, $T_2$, can be expressed as

$$T_2 = \frac{L}{1000f}$$

where $L$ is the optimal packet length in bit and $f$ is the optimal frequency in kHz. From Fig. 7, we find that $T_1 \gg T_2$. Thus, the assumption is validated.

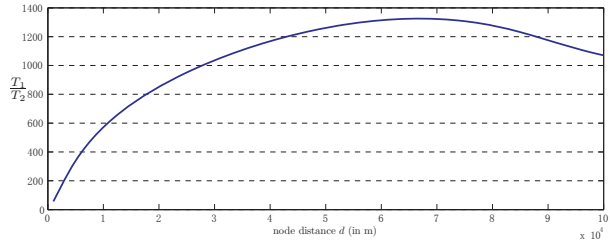

Figure 7: $\frac{T_1}{T_2}$ vs. $d$. For different $P_{acc0}$, the shape of the curve is almost the same.

## 5.4 Summary

In this section, we demonstrate that (30) and (32) constitute an accurate approximate solution to (22) given a wide range of parameters. We also show that (30) and (32) can find near optimum of (11).

We vary $d$ and $P_{acc0}$, calculate the corresponding optimum value of $\ln E'_b$ using the explicit approximation (30) and (32), and obtain Fig. 8.

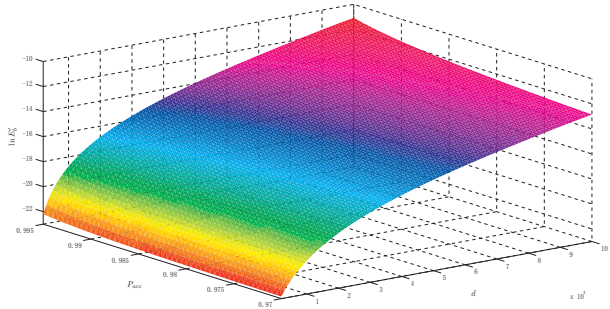

Figure 8: The optimum of (22).

From Fig. 6 and 8, we can see that the energy consumption grows as $P_{acc0}$ and $d$ become larger. This is consistent with intuition. When the node distance becomes larger, more energy is required keep a reliable transmission. On the other hand, more energy is necessary to meet higher $P_{acc0}$.

From Subsection 5.2, we know that $h_1$ is an effective constraint for the optimal solution to (22). This indicates that reliability and communication delay are crucial factors to the energy consumption for transmission. That is, concerning reliability and delay is a necessity when we intend to minimize energy consumption for transmission.

## 6  Conclusion

Following a rigorous and theoretical way, we address the issue of energy efficient transmission in UWSNs in this paper. Firstly we formalize an optimization problem which aims to minimize energy consumption but also comprehensively considers other performance metrics such as data reliability and communication delay. Using this problem, we study the effect of four key parameters (node distance, communication frequency, packet length and SNR) on the performance of UWSNs. Then we derive an explicit approximate solution to the optimization problem according to KKT conditions. This solution can provide theoretical guidelines for designing durable and reliable UWSNs and we show that the approximate solution is accurate with regard to a wide range of parameters. Our result also shows that reliability and communication delay are crucial factors to the energy consumption for transmission.